\documentclass[letter, traditabstract]{aa} 
\usepackage{multirow}
\usepackage{graphicx}
\usepackage{txfonts}
\begin{document}
   \title{A new $\alpha$-enhanced super-solar metallicity population}

   \author{V. Zh. Adibekyan\inst{1}\and  N. C. Santos\inst{1,2}\and  S.
G. Sousa\inst{1,3}\and  G. Israelian\inst{3,4}}

   \institute{Centro de Astrof\'{\i}ísica da Universidade do Porto, Rua das Estrelas,
4150-762 Porto, Portugal\\
\email{Vardan.Adibekyan@astro.up.pt}\and Departamento de F\'{\i}sica
e Astronomia, Faculdade de Ci\^{e}ncias da Universidade do Porto, Portugal\and Instituto
de Astrof\'{\i}sica de Canarias, 38200 La Laguna, Tenerife, Spain\and Departamento
de Astrof\'{\i}sica, Universidade de La Laguna, 38205 La Laguna, Tenerife,
Spain}

  \date{Received ...; Accepted...}

 
  \abstract
{We performed a uniform and detailed analysis of 1112 high-resolution
spectra of FGK dwarfs obtained with the HARPS spectrograph at the
ESO 3.6 m telescope (La Silla, Chile). Most stars have effective temperatures
4700 \emph{K} $\leq$ \emph{$T{}_{\mathrm{eff}}$} $\leq$ 6300 \emph{K
}and lie in the metallicity range of -1.39 $\leq$ {[}Fe/H{]} $\leq$
0.55. Our main goal is to investigate whether there are any
differences between the elemental abundance trends (especially {[}$\alpha$/Fe{]}
ratio) for stars of different subpopulations. The equivalent widths
of spectral lines are automatically measured from HARPS spectra with
the ARES code. The abundances of three $\alpha$ elements are determined
using a differential LTE analysis relative to the Sun, with the 2010
revised version of the spectral synthesis code MOOG and a grid of
Kurucz ATLAS9 atmospheres.The stars of our sample fall into two
populations, clearly separated in terms of {[}$\alpha$/Fe{]} up to super-solar
metallicities. In turn, high-$\alpha$ stars are also separated into
two families with a gap in both {[}$\alpha$/Fe{]} ( {[}$\alpha$/Fe{]}
$\approx$ 0.17) and metallicity ({[}Fe/H{]} $\approx$ -0.2) distributions.
The metal-poor high-$\alpha$ stars (thick disk) and metal-rich high-$\alpha$
stars are on average older than chemically defined\emph{ }thin disk
stars (low-$\alpha$ stars). The two $\alpha$-enhanced families have
different kinematics and orbital parameters. The metal-rich $\alpha$-enhanced
stars, such as thin disk stars have nearly circular orbits, close to the Galactic plane. We put forward the idea that
these stars may have been formed in the inner Galactic disk, but their
exact nature still remains to be clarified.}{}


   \maketitle
%

\section{Introduction}

   The investigation of stellar populations is very important to understand
the formation and evolution of our Galaxy. The Milky Way (MW) has
a composite structure with several subsystems. The main three stellar
populations of the MW in the solar neighborhood are the thin disk,
thick disk, and the halo. These populations have different kinematic
and chemical properties. The subdivision between the disk and halo was first
identified since long ago, but the thick disk was discovered far more recently
by Gilmore \& Reid (\cite{Gilmore}), who analysed the stellar
density distribution as a function of distance from the Galactic plane.

There is no obvious predetermined way to identify 
purely thick or thin disk stars in the solar neighborhood. There are
essentially three ways of distinguishing local thick and thin disk stars:
a purely kinematical approach (e.g. Bensby et al. \cite{Bensby-03},
\cite{Bensby-05}; Reddy et al. \cite{Reddy}), a purely chemical method (e.g.
Navarro et al. \cite{Navarro}), and by looking at a combination of
kinematics, metallicities, and stellar ages (e.g. Fuhrmann \cite{Fuhrmann};
Haywood \cite{Haywood-08a}).

The kinematic selection is a much more commonly applied method than the chemical
approach, because it is much easier to measure the velocity of a star
than to determine its chemical composition (particularly its $\alpha$-enhancement).
However, the chemical distinction of the disks can be more useful and reliable,
at least, because chemistry is a relatively more stable property of
a star, that is intimately connected to the time and place of its birth,
whereas spatial positions and kinematics are evolving properties.

During the past few years, there have been several studies directed to the detailed
elemental abundance investigations of stars in different subpopulations.
However, spectroscopic studies are in general limited to small samples
of a few hundred stars at most (e.g. Feltzing \& Gustafsson \cite{Feltzing & Gustafsson};
Bensby et al. \cite{Bensby-03,Bensby-05,Bensby-2007}; Soubiran et
al. \cite{Soubiran05}; Reddy et al. \cite{Reddy}; Ram\'{\i}rez
et al. \cite{Ram=0000EDrez}) and only a few studies have been based on samples as large  as 1000 stars 
(e.g. Gazzano et al. \cite{Gazzano-10}; Gazzano \cite{Gazzano-11}; Petigura \& Marcy \cite{Petigura-11}).
To investigate {[}$\alpha$/Fe{]} abundances
in the thin and thick disks with relatively large samples, some studies
combine data from different sources (e.g. Navarro et al. \cite{Navarro}).{{}
Alternatively, some other studies estimated the {[}$\alpha$/Fe{]} ratio
from Str\"{o}mgren indices (e.g. Casagrande et al. \cite{Casagrande})}.
However, both methods are far less precise than those obtained with high-resolution
spectroscopy, and prevent us from seeing any separation gap between the
thin and thick disks. To minimize any type of external and internal
``errors'', one needs to have as large and as homogeneous a sample as
possible, with reliable measurements of their chemical and kinematic
features. 

In this Letter, we investigate the possible differences in the elemental
abundance trends for stars of different subpopulations, using a stellar
sample of 1112 long-lived dwarf stars. To separate and investigate the
different Galactic stellar subsystems, we focus
on the {[}$\alpha$/Fe{]} ratio (here ``$\alpha$'' refers to the
average abundance of Mg, Si, and Ti). The extensive and full investigation
of this sample, will be more focused on the abundance difference between stars
with and without planets and be presented in an upcoming paper where
we will also describe the observations, data reductions, and abundance
analysis in detail.


\section{The sample}

The sample used in this work consists of 1112 FGK stars observed within
the context of the HARPS GTO programs, hereafter called HARPS-1 (Mayor
et al. \cite{Mayor}), HARPS-2 (Lo Curto et al. \cite{Lo Curto}),
and HARPS-4 (Santos et al. \cite{Santos_11}). The stars are slowly-rotating,
non-evolved, and in general have a low level of activity. The individual
spectra of each star were reduced using the HARPS pipeline and then
combined with IRAF%
\footnote{IRAF is distributed by National Optical Astronomy Observatories, operated
by the Association of Universities for Research in Astronomy, Inc.,
under contract with the National Science Foundation, U.S.A.%
} after correcting for its radial velocity. The final spectra have
a resolution of \emph{R }$\sim$110000 and signal-to-noise 
ratio (\emph{S/N}) ranging from $\sim$20 to $\sim$2000, depending on the amount
and quality of the original spectra. Fifty-five percent of the spectra
have \emph{S/N} higher than 200.

Precise stellar parameters for all the stars were determined in the
same manner and by the same authors from the same spectra used in
our study. For details, we refer to Sousa et al. (\cite{Sousa_08,Sousa_11,Sousa_11b}).
The typical uncertainties in the atmospheric parameters are of the
order of 30 \emph{K }for \emph{$T{}_{\mathrm{eff}}$} , 0.06 dex for
$\log\, g$, and 0.03 dex for {[}Fe/H{]}.

The stars in the sample have effective temperatures 4487 \emph{K}
$\leq$ \emph{$T{}_{\mathrm{eff}}$} $\leq$ 7212 \emph{K} (there
are only 12 stars with \emph{$T{}_{\mathrm{eff}}$} $>$ 6500 \emph{K)}
and metallicites -1.39 $\leq$ {[}Fe/H{]} $\leq$ 0.55 (only 11 stars
with {[}Fe/H{]} $<$ -1 and three stars with {[}Fe/H{]} $>$ 0.4), and they have surface gravities
2.68 $\leq$ $\log\,g$ $\leq$ 4.96 dex (again the number of ``outliers'' is very small, only 5 stars with $\log\, g$ $<$ 3.8 dex).


\section{\textbf{$\alpha$ element abundance}}

Elemental abundances for three $\alpha$ elements (Mg, Si, and Ti) were determined using a differential LTE analysis relative
to the Sun (the reference abundances were taken from Anders \& Grevesse (\cite{Anders-89}))
 with the 2010 revised version of the spectral synthesis code MOOG%
\footnote{http://www.as.utexas.edu/\textasciitilde{}chris/moog.html%
} (Sneden \cite{Sneden}) and a grid of Kurucz ATLAS9 atmospheres (Kurucz
et al. \cite{Kurucz}). The method and the atomic data are the same
as in Neves et al. (\cite{Neves}) (see also Adibekyan et al. \cite{Adibekyan},
in preparation). The equivalent widths (EW) were automatically measured
with the ARES%
\footnote{http://www.astro.up.pt/sousasag/ares%
} code (Automatic Routine for line Equivalent widths in stellar Spectra
- Sousa et al. \cite{Sousa_07}). The input parameters for ARES were
calculated following the procedure discussed in Sousa et al. (\cite{Sousa_11}).

The total uncertainties in the abundances were evaluated as combinations of the line-to-line scatter errors 
and errors induced by uncertainties in the model atmosphere parameters.The line-to-line scatter errors were 
estimated as $\sigma$/$\sqrt{N}$, where $\sigma$ is the standard deviation of \emph{N} measurements. The 
abundances of the elements were determined using 16, 3, 24, and 6 lines for Si, Mg, TiI, and TiII, respectively.
The average error in the {[}$\alpha$/Fe{]} ratio is 0.05 dex.
We note that, in general, our derived abundances agree with the {[}$\alpha$/Fe{]} presented in the literature 
(see Neves et al. \cite{Neves}; Adibekyan et al. \cite{Adibekyan}, in preparation).
 
Studying the [$\alpha$/Fe] error distribution with stellar parameters, 
we observe no discernible trends except in the case of the \emph{$T{}_{\mathrm{eff}}$} where we found that  
for cooler stars (\emph{$T{}_{\mathrm{eff}}$} $\lesssim$ 4900 K) the error increases, reaching about 0.1 dex for stars with 
\emph{$T{}_{\mathrm{eff}}$} $\cong$ 4500 K.
We also observe a trend with \emph{$T{}_{\mathrm{eff}}$} for [TiI/Fe] ratio and
we decided to establish a cutoff temperature - \emph{$T{}_{\mathrm{cutoff}}$}=4900 K (for details see Neves et al. \cite{Neves}).
The number of stars in the sample with \emph{$T{}_{\mathrm{eff}}$} $>$ 4900 K is 940.

\begin{figure}
\centering
\includegraphics[angle=270,width=0.8\linewidth]{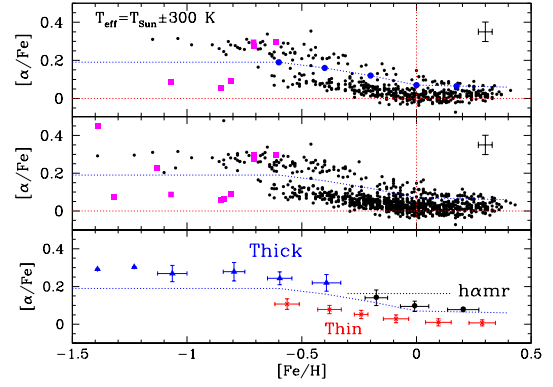}\caption{{[}$\alpha$/Fe{]} versus {[}Fe/H{]} for the whole sample (middle)
and for stars with \emph{$T{}_{\mathrm{eff}}$} =\emph{$T{}_{\mathrm{\odot}}$}
$\pm$300 K (top). Magenta squares refer to halo stars. Blue filled
circles are the separation points between low- and high-$\alpha$
stars, which are minimas of the {[}$\alpha$/Fe{]} histograms for
five metallicity bins (from {[}Fe/H{]} = -0.7 to 0.25) and the blue
dashed curve is the corresponding separation curve passing on that
points. The bottom panel is the {[}$\alpha$/Fe{]} versus {[}Fe/H{]}
plot for the whole sample in several bins of metallicity. The blue
triangles refer to the thick disk stars, black
filled circles to the metal-rich high-$\alpha$ stars (h$\alpha$mr), and the
red crosses to the thin disk stars. Error bars in the lower panel correspond
to the standard deviations, and the error bar in the upper and middle panels are the average errors in the {[}$\alpha$/Fe{]} and [Fe/H]. }
\end{figure}

\begin{figure}
\centering
\includegraphics[angle=270,width=0.7\linewidth]{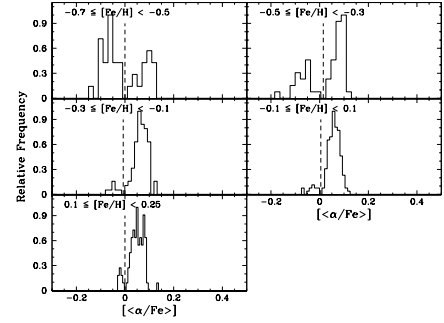}\caption{High-$\alpha$ and low-$\alpha$ separation histograms 
for the stars
with \emph{$T{}_{\mathrm{eff}}$} =\emph{$T{}_{\mathrm{\odot}}$}
$\pm$300 K after subtracting the separation curve. }
\end{figure}

Fig. 1 shows {[}$\alpha$/Fe{]} versus {[}Fe/H{]} for all stars in
the sample and for the stars with effective temperatures close to
the Sun by $\pm$300 K (the number of these stars is 483). As can be seen, the stars are clearly separated
into two groups according to the content of $\alpha$ elements: the
``high-$\alpha$'' and the ``low-$\alpha$'' stars (thin disk\footnote{We emphasize that the traditional selection 
(and therefore definitions) of thin and thick disk stars was based on their kinematics.}). 
This separation highlights the well-known $\alpha$
enhancement of thick disk stars relative to the thin disk found for
stars with {[}Fe/H{]} $<$ 0 (e.g. Fuhrmann \cite{Fuhrmann}; Bensby
et al. \cite{Bensby-03,Bensby-05}). The blue filled circles in Fig.
1 are the separation points between low- and high-$\alpha$ stars.
A separation into these two ``populations'' was performed based on 
{[}$\alpha$/Fe{]}, for the stars with \emph{$T{}_{\mathrm{eff}}$}
=\emph{$T{}_{\mathrm{\odot}}$} $\pm$300 K. We divided the sample
into five metallicity bins from {[}Fe/H{]} = -0.7 to 0.25 (see Fig.
2. for bin sizes) and identified the minima in the {[}$\alpha$/Fe{]}
histograms for each bin. The separation curve in Fig. 1 is the simple
connection of the above-mentioned separation points. In the metallicity
region {[}Fe/H{]} $<$ -0.7, all the stars, except five halo stars, are
$\alpha$-enhanced and lie above the {[}$\alpha$/Fe{]} = 0.19 line.
A distinction between low-$\alpha$ and high-$\alpha$ stars in the
supersolar metallicity region ({[}Fe/H{]} $>$ 0.25) was made using the
extrapolation of the separation line constructed from the last two
separation points. After subtracting the separation curve, we again
compiled the separation histograms presented in Fig. 2. As one
can see all the separation minimums are very close to {[}$\alpha$/Fe{]}
= 0, which proves that the separation is clear.

\begin{figure}
\centering
\includegraphics[angle=270,width=0.7\linewidth]{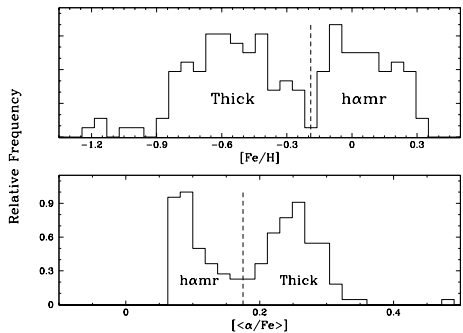}

\caption{The {[}Fe/H{]} and {[}$\alpha$/Fe{]} separation histograms for the
$\alpha$-enhanced stars.}
\end{figure}

It is interesting to see that high-$\alpha$ stars are also divided
into two subgroups: high-$\alpha$ metal-rich stars (hereafter ``h$\alpha$mr''), 
and high-$\alpha$ metal-poor stars (thick disk).
Fig. 3 shows the {[}Fe/H{]} and {[}$\alpha$/Fe{]} separation histograms
for the $\alpha$-enhanced stars. In this figure, one can see that
there is a gap between these two groups in terms of both  the metallicity (Fe/H
$\sim$ -0.2) and {[}$\alpha$/Fe{]} ({[}$\alpha$/Fe{]} $\sim$
0.17 ). This gap (in {[}$\alpha$/Fe{]}) can also be seen
 in Reddy et al. (\cite{Reddy}) (Fig. 20), although they 
differentiated between thin and thick disk stars using their kinematics.

On the basis of the above discussed histograms, we can define the metallicity
ranges for each population. We find that for thick disk stars, 
metallicities vary from -1.4 to -0.2 dex. For thin disk stars, 
the metallicities range roughly from  -0.7 to +0.4 dex.
The h$\alpha$mr group stars lie in the metallicities range 
roughly from -0.2 to +0.3 dex. These values, in particular the upper limits
to the thick disk and h$\alpha$mr, and the lower limits to the thin and h$\alpha$mr groups, may represent
the metallicity boundaries of the different populations. 

\begin{figure}
\centering
\includegraphics[angle=270,width=0.65\linewidth]{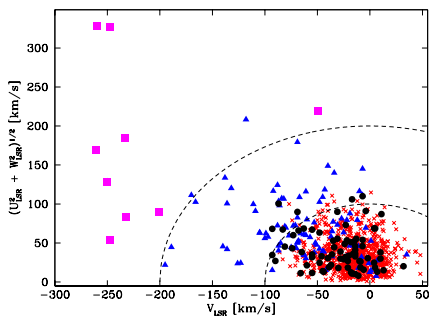}\caption{The Toomre diagram for the entire sample. 
The blue triangles and red crosses refer to the chemically selected thick and thin disk stars,
 and the black filled circles are h$\alpha$mr stars. Magenta squares refer to halo stars selected by their kinematics. }
\end{figure}

The magenta squares in Fig. 1 refer to stars belonging to halo (selected
by their kinematics, see Section 4). As is clearly evident, halo stars can
also be divided into two high-$\alpha$ and low-$\alpha$ groups. This
dichotomy confirms the results provided in Nissen \& Schuster (\cite{Nissen}),
where they present clear evidence of two distinct halo populations.

\section{Kinematics}

To study the kinematic relations and differences between
the three above separated groups, we computed their Galactic velocities.
The Galactic space velocity components (UVW) were derived with respect
to the local standard of rest (LSR), adopting the standard solar motion
(U$_{\odot}$, V$_{\odot}$, W$_{\odot}$) = (10.00, 5.25, 7.17) km
$\mathrm{s}^{-1}$ of Dehnen \& Binney (\cite{Dehnen & Binney}).
The main source of the parallaxes and proper motions were the updated
version of the Hipparcos catalog (van Leeuwen \cite{van Leeuwen}) and
the radial velocities obtained from the HARPS spectra. The parallaxes
with errors larger than 10\%, (which is true for less than 5\% of the stars in the sample) were redetermined following the procedure
described in Sousa et al. (\cite{Sousa_11}).
The number of stars with inaccurate proper motions (errors larger than 10\%) is less than 8\%. 
We did not perform a quality selection of them, because these errors in general do not 
change their memberships of a certain population. 
Combining the measurement errors in the parallaxes,
proper motions, and radial velocities, the resulting average errors in the U, V, and W velocities
are 2, 2.3, and 1.5 km \textit{s}$^{-1}$, respectively.

\begin{table}
\centering
\caption{The number of stars in different populations.}
\begin{tabular}{llcccc}
\hline
\hline 
 &  & \multicolumn{2}{c}{{\tiny Entire sample}} & \multicolumn{2}{c}{\emph{\tiny $T{}_{\mathrm{eff}}$}{\tiny{}
=}\emph{\tiny $T{}_{\mathrm{\odot}}$}{\tiny{} $\pm$300 K}}\tabularnewline
 &  & {\tiny Bensby} & {\tiny Robin} & {\tiny Bensby} & {\tiny Robin}\tabularnewline
\hline 
\multirow{3}{*}{{\tiny Thin}} & {\tiny Thin} & {\tiny 699} & {\tiny 742} & {\tiny 353} & {\tiny 376}\tabularnewline
 & {\tiny Thick} & {\tiny 17} & {\tiny 5} & {\tiny 13} & {\tiny 4}\tabularnewline
 & {\tiny Transition} & {\tiny 46} & {\tiny 16} & {\tiny 25} & {\tiny 11}\tabularnewline
\cline{2-6} 
\multirow{3}{*}{{\tiny Thick}} & {\tiny Thin} & {\tiny 22} & {\tiny 33} & {\tiny 13} & {\tiny 20}\tabularnewline
 & {\tiny Thick} & {\tiny 54} & {\tiny 42} & {\tiny 33} & {\tiny 27}\tabularnewline
 & {\tiny Transition} & {\tiny 17} & {\tiny 18} & {\tiny 11} & {\tiny 10}\tabularnewline
\cline{2-6} 
\multirow{3}{*}{{\tiny h$\alpha$mr}} & {\tiny Thin} & {\tiny 61} & {\tiny 64} & {\tiny 25} & {\tiny 26}\tabularnewline
 & {\tiny Thick} & {\tiny 10} & {\tiny 3} & {\tiny 5} & {\tiny 3}\tabularnewline
 & {\tiny Transition} & {\tiny 4} & {\tiny 8} & {\tiny 0} & {\tiny 1}\tabularnewline
\hline 
\end{tabular}
\end{table}

We adopted the method of assigning the probability of each star to
either the thin disk, the thick disk, or the halo described in Reddy
et al. (\cite{Reddy}). This assumes that the sample is a mixture of
the three populations and each population follows a Gaussian distribution
of random velocities in each component (Schwarzschild \cite{Schwarzschild}).
In this paper, we adopt the mean values (asymmetric drift) and dispersion
in the Gaussian distribution, and the population fractions were taken
from Bensby et al. (\cite{Bensby-03}) and Robin et al. (\cite{Robin}).
We consider that a probability in excess of 70\% suffices to assign
a star to the concrete population.

\begin{table*}
\centering
\caption{The average values of the ages, eccentricities, and Z$_{\mathrm{max}}$
for the three stellar groups, along with their \emph{rms} and the
number of stars.}
\begin{tabular}[t]{lllll}
\hline
\hline 
{\tiny Stellar groups} & {\tiny BASTI age} & {\tiny Padova age} & {\tiny Eccentricity} & {\tiny Z$_{\mathrm{max}}$}\tabularnewline
\hline 
{\tiny Thin disk} & {\tiny 5.4$\pm$2.2 (555)} & {\tiny 4.8$\pm$2.0 (554)} & {\tiny 0.14$\pm$0.1 (549)} & {\tiny 0.25$\pm$0.25 (546)}\tabularnewline
{\tiny Thin disk stars with }\emph{\tiny $T{}_{\mathrm{eff}}$}{\tiny{}
=}\emph{\tiny $T{}_{\mathrm{\odot}}$}{\tiny{} $\pm$300 K} & {\tiny 5.6$\pm$2.2 (313)} & {\tiny 4.9$\pm$1.9 (313)} & {\tiny 0.15$\pm$0.11 (314)} & {\tiny 0.27$\pm$0.29 (311)}\tabularnewline
{\tiny Thick disk} & {\tiny 8.3$\pm$2.5 (84)} & {\tiny 8.1$\pm$2.6 (84)} & {\tiny 0.31$\pm$0.16 (75)} & {\tiny 1.01$\pm$1.31 (83)}\tabularnewline
{\tiny Thick disk stars with }\emph{\tiny $T{}_{\mathrm{eff}}$}{\tiny{}
=}\emph{\tiny $T{}_{\mathrm{\odot}}$}{\tiny{} $\pm$300 K} & {\tiny 8.7$\pm$2.3 (56)} & {\tiny 8.5$\pm$2.4 (56)} & {\tiny 0.31$\pm$0.17 (51)} & {\tiny 1.17$\pm$1.5 (56)}\tabularnewline
{\tiny h$\alpha$mr stars} & {\tiny 8.0$\pm$3.0 (45)} & {\tiny 7.1$\pm$2.7 (45)} & {\tiny 0.18$\pm$0.1 (45)} & {\tiny 0.32$\pm$0.34 (45)}\tabularnewline
{\tiny h$\alpha$mr stars with }\emph{\tiny $T{}_{\mathrm{eff}}$}{\tiny{}
=}\emph{\tiny $T{}_{\mathrm{\odot}}$}{\tiny{} $\pm$300 K} & {\tiny 9.6$\pm$2.4 (24)} & {\tiny 8.4$\pm$2.2 (24)} & {\tiny 0.2$\pm$0.1 (26)} & {\tiny 0.33$\pm$0.41 (28)}\tabularnewline
\hline 
\end{tabular}
\end{table*}

These kinematic separation criteria suggest that most of the stars
in the h$\alpha$mr stellar family, such as chemically defined thin disk stars, have thin disk kinematics
(see Table 1). The same impression is created by Fig. 4, where
all the stars in this study (with \emph{$T{}_{\mathrm{eff}}$} $>$ 4900 K) are shown in a Toomre diagram. The chemically defined ``thick'' disk stars
mainly have thick disk kinematics.

\section{Discussion}

The near constancy and the high level of $\alpha$-element abundances
relative to Fe for the metal-poor thick disk stars ({[}Fe/H{]} $\lesssim$
-0.3) suggest that they formed in regions of high star formation
rate where massive stars enriching the interstellar medium with $\alpha$-elements
explode as core-collapse supernovae (SNe) (see, e.g. Ballero et
al. 2007b). Many papers report the existence of the \textquotedblleft{}knee\textquotedblright{}
in {[}$\alpha$/Fe{]} trends for the thick disk stars (kinematically
selected) when {[}Fe/H{]} reaches to $\approx$ -0.3 (see, e.g. Feltzing
et al. \cite{Feltzing}; Bensby et al. \cite{Bensby-2007}). This
downturn in {[}$\alpha$/Fe{]}, up to solar metallicities (Bensby et
al. \cite{Bensby-2007}), means that low- and intermediate-mass stars
start to play a significant role in the chemical enrichment, through
the explosions of SN Ia, which produce relatively little of the $\alpha$-elements.
This downturn is seen in Fig 1. As already noted, however, there is
a gap between metal-poor and metal-rich $\alpha$-enhanced stars,
which casts doubt on the relation between the h$\alpha$mr and the thick disk stars.

To study the differences and/or the similarities of the Galactic
orbital parameters and ages between two $\alpha$-enhanced groups,
we cross-matched our sample with the Geneva-Copenhagen Survey (GCS)
sample (Casagrande et al. \cite{Casagrande}), which provides the
eccentricities of the orbits, a maximum vertical distance (Z$_{\mathrm{max}}$)
a star can reach above the Galactic plane, and the ages of about 700
of our stars.

\begin{figure}
\centering
\includegraphics[angle=270,width=0.65\linewidth]{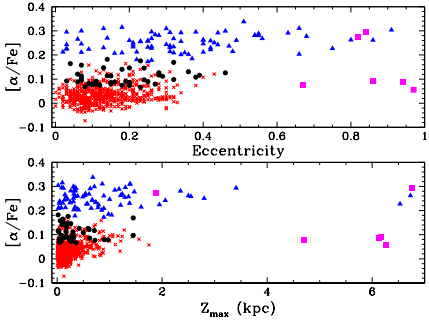}\caption{{[}$\alpha$/Fe{]} versus eccentricity and Z$_{max}$. 
The symbols are the same as in Fig. 4. }
\end{figure}

The mean ages of the three groups and their standard deviations are
presented in Table 2. As can be seen from the table, the thick disk and
h$\alpha$mr family stars have almost the same age, being on average older
than thin disk stars by about 3 Gyr.

Our study of the orbital parameters has uncovered a different connection between
the three groups. Most of the h$\alpha$mr stars, such as the thin disk
stars, have nearly circular orbits, close to the Galactic plane.
Unlike them,  thick disk stars move in more
eccentric orbits and have on average vertical distances three times
larger than other two groups. 
The [$\alpha$/Fe] distributions as a function of eccentricities and Z$_{max}$ are shown in Fig. 5.
 The mean values of the eccentricities
and Z$_{max}$ for the above separated three stellar families are
presented in Table 2.

We can summarize the results obtained in this letter in the following way: our
sample of stars, which consists of 1112 FGK long-lived dwarfs, can be
clearly separated into two groups according to the content of $\alpha$
elements. In turn, high-$\alpha$ stars can also be divided
into two families based on the gap between more $\alpha$-enhanced
(thick disk) and less $\alpha$-enhanced (h$\alpha$mr) stars.
Studying these three stellar families, we see that h$\alpha$mr stars have
orbits similar to the thin disk stars, but that they are similar
to thick disk stars in terms of age.This unusual results for metal-rich stars
assigned to both the thin and thick disks (by their kinematics) 
was found previously in Soubiran et al. (\cite{Soubiran05}).

In the past couple of years, many studies have focused
on the abundance determination of the bulge and inner disk stars (Fulbright
et al. \cite{Fulbright-05,Fulbright}; Mel\'{e}ndez et al. \cite{Mel=0000E9ndez};
Alves-Brito et al. \cite{Alves-Brito}; Epstein et al. \cite{Epstein};
Bensby et al. \cite{Bensby-2010a,Bensby-2010b,Bensby-2011,Bensby-2011a}).
The abundance studies for the bulge dwarfs at sub-solar metallicities
show excellent agreement with the abundance patterns in the local
thick disk (Mel\'{e}ndez et al. \cite{Mel=0000E9ndez}; Alves-Brito et
al. \cite{Alves-Brito}; Bensby et al. 2010\cite{Bensby-2010b}).{{}
This suggests that these two populations have comparable chemical histories.
It is also intriguing that at super-solar metallicities bulge stars
seems to be more $\alpha$-enhanced than local thin disk stars
(Fulbright et al. \cite{Fulbright-05,Fulbright}, see Fig. 10 in Bensby
et al. 2010\cite{Bensby-2010b} and Fig. 10 in Alves-Brito et al.
\cite{Alves-Brito}). Taking into account that the high-$\alpha$
metal-rich stars of our sample are on average as old as bulge stars,
and adding to this the aforementioned similarities between the different
populations, we can see a link between the h$\alpha$mr stars 
and stars in the inner disk. Stellar radial migration could give an explanation
of this link (e.g. Haywood \cite{Haywood-08b}; Sch\"{o}nrich  \& Binney  \cite{Schonrich-09a}). 
It has been proposed that metal-rich stars found
in the solar vicinity may have been formed in the inner Galactic disk
regions (e.g. Grenon \cite{Grenon-99}; Ecuvillon et al. \cite{Ecuvillon-07};
Famaey et al. \cite{Famaey-07}; Santos et al. \cite{Santos-08}; Sch\"{o}nrich  \& Binney  \cite{Schonrich-09b}). 

Nevertheless, the origin and nature of these stars remains unclear
and needs to be clarified. Although the present observations suggest
that h$\alpha$mr stars (high-alpha, metal rich) may have originated from
the inner disk (e.g. inner thick-disk members), they do not allow
us to exclude the possibility that they represent a whole new Galactic
population. More observations are needed to resolve this uncertainty. 

\begin{acknowledgements}
V.Zh.A. and S.G.S. acknowledge the support from the Fundação para a
Ciência e Tecnologia (FCT) in the form of a grants SFRH/BPD/70574/2010
and SFRH/BPD/47611/2008, respectively. N.C.S. acknowledges the support
of the European Research Council/European Community under the FP7
through a Starting Grant, as well as the support from (FCT), Portugal,
through program Ciência 2007. {We also acknowledge
support from FCT in the form of grant reference PTDC/CTE-AST/098528/2008. }\end{acknowledgements}

\end{document}